\documentclass[iop]{emulateapj}

\newcommand{\mearth}{M$_{\oplus}$}

\slugcomment{Accepted to the Astrophysical Journal}

\shorttitle{MAGNETODYNAMOS IN ROCKY EXOPLANETS}
\shortauthors{GAIDOS ET AL.}

\begin{document}


\title{Thermodynamic Limits on Magnetodynamos in Rocky Exoplanets}


\author{Eric Gaidos and Clinton P. Conrad}
\affil{Department of Geology and Geophysics, University of Hawaii at Manoa, Honolulu, HI 96822}

\and

\author{Michael Manga\altaffilmark{1} and John Hernlund}
\affil{Department of Earth and Planetary Science,
University of California at Berkeley, Berkeley, California, USA 94720.}


\altaffiltext{1}{Center for Integrative Planetary Science, University
of California, Berkeley}


\begin{abstract}
To ascertain whether magnetic dynamos operate in rocky exoplanets more
massive or hotter than the Earth, we developed a parametric model of a
differentiated rocky planet and its thermal evolution.  Our model
reproduces the established properties of Earth's interior and magnetic
field at the present time.  When applied to Venus, assuming that
planet lacks plate tectonics and has a dehydrated mantle with an
elevated viscosity, the model shows that the dynamo shuts down or
never operated.  Our model predicts that at a fixed planet mass,
dynamo history is sensitive to core size, but not to the initial
inventory of long-lived, heat-producing radionuclides.  It predicts
that rocky planets larger than 2.5 Earth masses will not develop inner
cores because the temperature-pressure slope of the iron solidus
becomes flatter than that of the core adiabat.  Instead, iron ``snow''
will condense near or at the top of these cores, and the net transfer
of latent heat upwards will suppress convection and a dynamo.  More
massive planets can have anemic dynamos due to core cooling, but only
if they have mobile lids (plate tectonics).  The lifetime of these
dynamos is shorter with increasing planet mass but longer with higher
surface temperature.  Massive Venus-like planets with stagnant lids
and more viscous mantles will lack dynamos altogether.  We identify
two alternative sources of magnetic fields on rocky planets: eddy
currents induced in the hot or molten upper layers of planets on very
short period orbits, and dynamos in the ionic conducting layers of
``ocean'' planets with $\sim$10\% mass in an upper mantle of water
(ice).
\end{abstract}





\section{Introduction}

Earth distinguishes itself in the inner Solar System with a magnetic
dipole generated by a dynamo in its iron core.  Mars had a magnetic
field of comparable intensity that collapsed early in the planet's
history \citep{Arkani04} and never re-appeared \citep{Lillis08}.
Mercury has a weak (1\% of Earth) global field
\citep{Stanley05,Christensen06b}.  Venus currently lacks a dynamo, and
we have no evidence for or against one in its past \citep{Nimmo02}.
At least one massive satellite in the outer Solar System (Ganymede)
appears to have a core dynamo \citep{Schubert96}.\\

Planets as small as twice the mass of the Earth have been discovered
on short-period orbits around other stars \citep{Mayor09}.  A fraction
of these transit the parent star, allowing mean densities to be
determined.  The smallest transiting exoplanet has a mass of
$4.8\pm0.8$ \mearth~and a mean density of $5.6\pm 1.3$ g~cm$^{-3}$,
consistent with an Earth-like composition \citep{Leger09,Queloz09}.
The second smallest ($6.6\pm0.9$ \mearth) has a mean density only
one-third that of Earth, indicating it has a thick volatile envelope
\citep{Charbonneau09}. These two planets may rotate synchronously and
have effective emitting temperatures of around 2600~K and 500~K,
respectively (that of Earth is 255~K).  The {\it Kepler} spacecraft
mission \citep{Koch10,Borucki10} is expected to find many such ``hot''
massive rocky planets \citep{Selsis07,Gaidos07}.\\

Do these planets have magnetodynamos?  While there have been many
theoretical studies of the dynamos in the Earth and smaller Solar
System bodies, as well as in the ice and gas giants
\citep{Breuer09,Stevenson09}, rocky planets more massive than the
Earth have not been considered, although simple dynamo scaling laws
have been applied to gas giants on short-period orbits
\citep{Griessmeier04}.  Magnetic fields may protect planetary
atmospheres against erosion by stellar winds and coronal mass
ejections \citep{Griessmeier04,Lundkin07,Dehant07}.  Interaction
between a giant planet's magnetic field and that of the host star has
been inferred from corotational chromospheric activity
\citep{Shkolnik08,Walker08} and may produce detectable radio emission
\citep{Lecavelier09}.  Ohmic dissipation of the kinetic energy of
winds in partially ionized atmospheres has been proposed as an
explanation for the ``inflated'' radii of some short-period gas giants
\citep{Batygin10}.\\

For a planet to have a dynamo, it must contain a fluid layer that is
sufficiently electrically conducting for the magnetic Reynolds number
Re$_m$ = $V L/\lambda > 40$, where $V$ and $L$ are a characteristic
fluid velocity and length, respectively, $\lambda = 1/(\mu_0 \sigma$)
is the magnetic diffusivity, $\mu_0 = 4\pi \times 10^{-7}$~N~A$^{-2}$,
and $\sigma$ is the electrical conductivity.  In Earth-like planets
the conducting fluid is liquid iron or an alloy thereof ($\lambda \sim
2$ m$^2$ sec$^{-1}$ \citet{Stevenson09}).  The possibility that the
oxidation state of planetary material might prevent iron core
formation was explored by \citet{Elkins08}, but we assume that such
cores do form.  Metallization of silicates and miscibility with Fe is
not expected until a pressure of 20~TPa \citep{Oganov05}.  In the
cores of Earth-size planets ($L \sim 3 \times 10^6$ m), convective
motions as small as $10^{-4}$~m~sec$^{-1}$ are sufficient for Re$_m$
$> 40$.  The temporal variation of Earth's magnetic field, such as
westward drift, imply velocities of $10^{-4}-10^{-3}$~m~sec$^{-1}$ at
the top of the core.\\

A dynamo also requires a source of energy to maintain convective
motions against internal ohmic dissipation.  Core convection can be
driven by the removal of heat from the core faster than it can be
transported by conduction, by the release of latent heat during phase
transitions (i.e., solidification of iron), or by the formation of
buoyant fluid via the expulsion of light elements during iron
solidification.  The last two are related to the first because the
latent heat of solidification must be removed.  Thus dynamo activity
is ultimately related to the transport of heat and the temperature
contrast across the core-mantle boundary (CMB).  That contrast can be
maintained either by efficient cooling of the lower mantle (e.g., by
descending slabs of former lithosphere during plate tectonics) or by
the heating of the core by radioactive nuclides.  The absence of a
dynamo in Venus has been ascribed to the lack of plate tectonics,
which slows cooling of the mantle, as well as the core
\citep{Nimmo02}.\\

\begin{figure}
\noindent\includegraphics[width=3in]{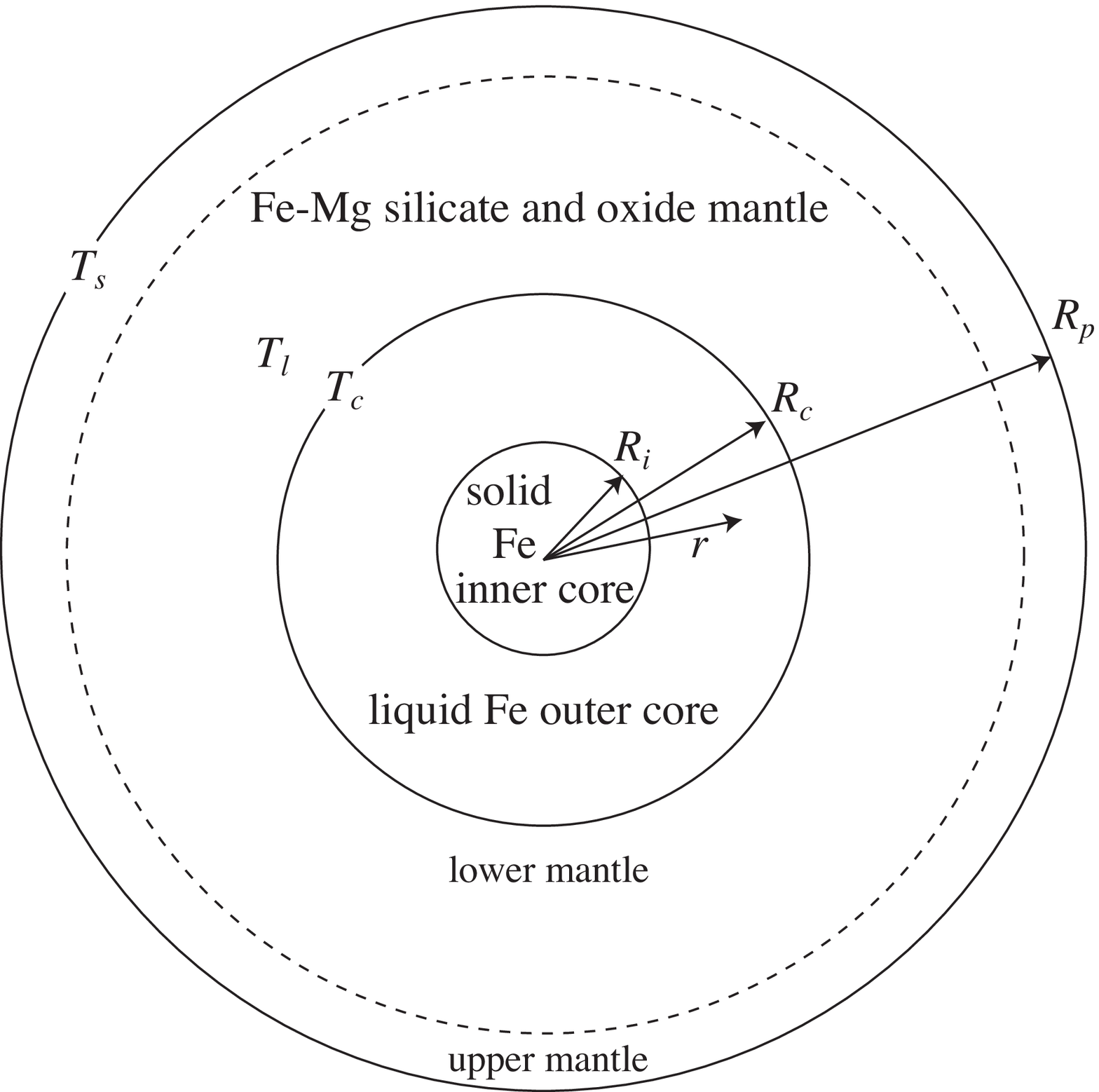}
\caption{Schematic of our idealized planet.}
\label{fig.schematic}
\end{figure}

In dynamo theory, the condition for convection is expressed as a
requirement on entropy production, rather than energy production,
because the latter is unaffected by dynamo dissipation when mechanical
energy is converted to heat.  The balance between entropy sinks and
sources in the core gives the maximum dissipation $\Phi$ by the
dynamo.  If $\Phi > 0$ then a dynamo is permitted.  We calculate
$\Phi$ using parameterized models of the interior structure and
thermal evolution of a planet consisting of an Fe core and a silicate
mantle.  We account for, but do not vary, the presence of light
elements in the core.  We estimate the magnetic Reynolds number and
average surface field using previously established scaling laws.  We
address whether dynamos operate in rocky planets with different
masses, surface temperatures, and modes of mantle convection.  Figure
\ref{fig.schematic} is a schematic of our simplified planet.  In the
equations that follow, we employ the following subscripts: 1
(ambient), p (planet or surface), m (mantle), c (core or core-mantle
boundary), i (inner core boundary), and 0 (center).\\

\section{Insight from scaling laws}

The manetostrophic approximation assumes that Lorentz forces are
balanced by Coriolis forces in the core \citep{Stevenson03}.  This
produces a scaling relationship between magnetic field strength (or
dipole moment) and core mass, radius, and rotation rate.
Alternatively, ohmic dissipation, and hence the strength of the
dynamo, are set by the available power \citep{Christensen06}.
Analysis of numerous numerical dynamo simulations shows that, in this
regime, the RMS strength of the field in the core, $B_c$, can be
related to a Rayleigh number based on a mass anomaly flow (mass
deficit advected per unit time)
\citep{Christensen06,Aubert09,Christensen09}, and through that to the
convective power in the core \citep{Buffett96}.  In the case where the
magnetic Prandtl number is much less than one and essentially all the
convective power is lost to ohmic dissipation, that relationship can
be expressed in terms of a dimensionless convective power $p$
\begin{equation}
\label{eqn.bfield}
B_c = a_1 \sqrt{\mu_0 \bar{\rho}} \Omega (R_o - R_i) p^{b_1}
\end{equation}
where $a_1$ and $b_1$ are dimensionless parameters, $\bar{\rho}$ is
the mean density of the core, $\Omega$ is the angular rotation rate of
the planet, $R_o$ and $R_i$ are the outer and inner boundaries of the
convecting zone, and $p = \phi \bar{T}/\left[\Omega^3
(R_c-R_i)^2\right]$.  $\phi \bar{T}$ is the available convective power
per unit mass, where $\phi = \Phi/M_c$ is the entropy available per
unit mass and time, and $\bar{T}$ is an effective dissipation
temperature.  \citet{Aubert09} find $a_1 \approx 1.65$ and $b_1$ to be
almost exactly 1/3.  The latter eliminates the dependence on $\Omega$:
\begin{equation}
\label{eqn.bfield2}
B_c \approx a_1 \sqrt{\mu_0 \bar{\rho}}\left[\phi \bar{T}
(R_c - R_i)\right]^{1/3}.
\end{equation}
Likewise, the magnetic Reynolds number can be found by a relationship
between the Rossby number and $p$, giving:
\begin{equation}
\label{eqn.rem1}
{\rm Re}_m = a_2 \frac{\Omega (R_c-R_i)^2}{\lambda} p^{b_2},
\end{equation}
where again $a_2$ and $b_2$ are dimensionless.  $b_2$ is not
necessarily 1/3 and this introduces a slight dependence on $\Omega$.
For example, \citet{Aubert09} find $a_2 \approx 1.3$ and $b_2 \approx
0.42$.  For the purpose of a simple order-of-magnitude estimate of
Re$_m$, we ignore these complications, set $b_2 = 1/3$, producing a
new formulation:
\begin{equation}
\label{eqn.rem2}
{\rm Re}_m \approx a'_2 \frac{R_c - R_i}{\lambda}\left[\phi \bar{T}
\left(R_c-R_i\right)\right]^{1/3}
\end{equation}
For plausible terrestrial values ($\phi \sim 100$~MW~K$^{-1}$,
$\bar{T} \sim 5000$~K; \citet{Labrosse07}), Re$_m \sim a'_2 10^4$.  So
as long $a'_2 \sim 1$ its exact value will not be critical.  Combining
Equations \ref{eqn.bfield2} and \ref{eqn.rem2} for terrestrial values
gives Re$_m \sim 16 (B_c/1 \mu{\rm T})$. Thus if the predicted field
in the core is at least a few $\mu$T, i.e. the surface field
(attenuated at least by the cubic distance law of a dipole) is $\geq 1
\mu$T, then Re$_m > 40$.  This criterion is only very weakly dependent
on planet mass.

The strength of the field at the surface of the planet and beyond will
be sensitive to the field's topology, especially the fraction in the
dipole component, and will depend on $\Omega$.  More rapidly rotating
planets (lower Rossby number) will have a stronger dipole component
\citep{Christensen06}.  In the case of a pure dipole, the average
surface field will be
\begin{equation}
\label{eqn.bsurf}
B_p \approx B_c (R_c/R_p)^3.
\end{equation}
We estimate the $B_p$ and Re$_m$ using Equations \ref{eqn.bfield2},
\ref{eqn.rem2}, and \ref{eqn.bsurf}.  We conservatively take $\bar{T}$
to be the temperature at the top of the core.  We assume that the
field is dipole-dominated like the Earth.

\section{Model}

\subsection{Interior structure \label{sec.interior}}

A planet is modeled as a homogeneous, fully convecting Mg/Fe-silicate
mantle surrounding a liquid/solid Fe core.  The pressure-density
profiles are calculated self-consistently with the (time-varying)
size of the inner core.  Third-order Birch-Murnagham (BM) equations of
state (EOS) are used for each component.  The pressure is:
\begin{equation}
P = \frac{3}{2}K_1\left(x^7-x^5\right)\left[1 + \frac{3}{4}\left(4-K_1'\right)\left(1-x^2\right)\right],
\end{equation}
where $x = (\rho/\rho_1)^{1/3}$ is the isotropic strain.  Values of
the ambient density $\rho_1$, compressibility $K_1$ and its pressure
derivative $K_1'$ for the relevant planetary materials are given in
Table 1.  Electron degeneracy (Thomas-Fermi-Dirac, or TFD) pressure
will be important in the metal cores of massive Earth-like planets.
We calculated the TFD contribution to the pressure using the
formulation of \citet{Zapolsky69} and found that the EOS for liquid
and $\epsilon$-phase solid iron (Fe(l) and Fe($\epsilon$)) intersect
at about 3.3~TPa.  Pressures beyond 3~TPa are not relevant to the
range of masses considered here.  The density in the liquid part of
the core is adjusted by a fixed fraction $\delta \rho/\rho$ to account
for the presence of light elements \citep{Li03}.\\

\begin{figure}
\noindent\includegraphics[width=3in]{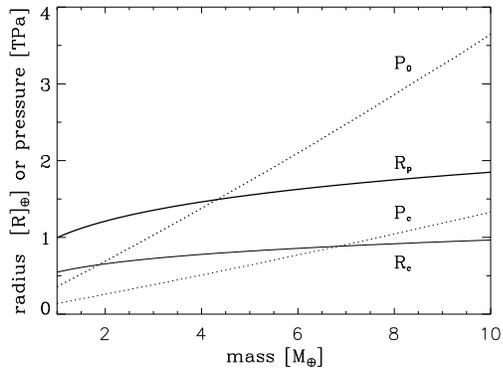}
\caption{Predicted radius of an Earth-like planet $R_p$, and its iron
core $R_c$, and the central and core-mantle-boundary pressures, $P_0$
and $P_c$, as a function of total mass.}
\label{fig.massradius}
\end{figure}

Our interior model reproduces the central pressure (364~GPa), CMB
pressure (136~GPa), and core size $R_c$ (3480~km) of the Earth
\citep{Dziewonski81}, but only if the zero-pressure compressibility
for Fe(l) is 40\% larger than the 106~GPa value reported by
\citet{Anderson94} based on shock experiments.  With this adjustment,
the predicted radius of the Earth is 25~km (0.4\%) less than the
actual value, but this is not surprising because we do not include low
density crustal rocks or an ocean.  Our dynamo predictions are much
more sensitive to the properties of the core than those of the entire
planet.  Predicted planet radius, core radius, pressure at the CMB,
and central pressure are plotted versus total mass in Figure
\ref{fig.massradius}.  Pressures increase as the core solidifies and
contracts.\\

\subsection{Temperature and density profiles \label{sec.tempprof}}

The adiabatic temperature profile in the outer convecting core is
approximately
\begin{equation}
\label{eqn.tempprofile}
T(r) \approx T_c \exp \left[(R_c^2 - r^2)/D_0^2\right],
\end{equation}
where the thermal length scale evaluated at the planet center is,
\begin{equation}
D_0 = \sqrt{\frac{3c_p}{2\pi \alpha_0 \rho_0 G}},
\end{equation}
\citep{Labrosse03}, $c_p$ and $\alpha$ are the heat capacity and
thermal expansivity of Fe, respectively, and $G$ is the gravitational
constant.  The temperature profile deviates from Equation
\ref{eqn.tempprofile} to the extent that $D$, and specifically
$\alpha$, vary with depth in the outer core.  $\alpha$ must therefore
be evaluated appropriately when calculating terms that are sensitive
to $D$, i.e. the heat and entropy transported by thermal conduction.
Hereafter $D$ (without a subscript) is $D_0$.\\

The density in the core follows
\begin{equation}
\label{eqn.densprofile}
\rho(r) \approx \rho_c \exp \left[(R_c^2 - r^2)/L^2\right],
\end{equation}
where the density scale length is
\begin{equation}
L = \sqrt{\frac{9K_1}{2\pi G \rho_1 \rho_0} \left( \ln \frac{\rho_0}{\rho_1} + 1 \right)}.
\end{equation}
For the Earth, $D \sim 6400$~km and $L \sim 7400$~km and these values
are only weakly dependent on planet size.  We expect the core radius
in planets of Earth-like composition to scale as $M_p^{0.25}$
\citep{Sotin07,Seager07}.  The $n$th order term in either $r/D$ or
$r/L$ will be of order $M_p^{n/4}/2^n$.  For planets of a few Earth
masses, we must retain terms in expansions with $r/D$ or $r/L$ through
fifth order.\\

If a solid inner core is present, the temperature at the inner-outer
core boundary is the intercept between the adiabat and the iron
melting point $\tau$.  Adopting a Lindemann law for $\tau$,
\begin{equation}
\label{eqn.lindeman}
\frac{\partial \log \tau}{\partial \log \rho} = 2\left(\gamma - \frac{1}{3}\right)
\end{equation}
where $\gamma$ is the Gr\"{u}neisen parameter.  The application of the
Lindemann law to the deep interiors of planets is sometimes viewed as
speculative (e.g., \citet{Wolf84}). The assumptions upon which it is
derived are not strictly valid for polyatomic systems, and the law is
known to sometimes fail for minerals with complex interatomic forces
and structures. However, at high pressures and for metals it becomes a
better approximation, and hence is widely used to interpret shock data
(e.g., \citet{Anderson96}). Importantly, it appears to provide a good
description of experimental data and first principle calculations for
Fe (e.g., \citet{Hemley01,Huser07}) and other metals
(e.g. \citet{Dai02}).

The radial dependence of the solidus becomes
\begin{equation}
\tau(r) = \tau_0 \exp \left[-2\left(1-\frac{1}{3\gamma}\right)\left(\frac{r}{D}\right)^2\right].
\end{equation}
where we have used $L^2/D^2 = \gamma$ \citep{Labrosse03}.  Thus, the
temperature at the top of the core is uniquely set by the ratio of the
inner to outer core radius $\Re = R_i/R_c$:
\begin{equation}
T_c = \tau_0 \exp \left[-\left[1 + \Re^2 \left(1 - \frac{2}{3\gamma}\right)\right]\frac{R_c^2}{D^2}\right].
\end{equation}
$T_c$ must decrease as the inner core grows.  If the entire core is
liquid, then $T_c$ is the independent variable.  For greater accuracy,
we use our interior model to directly calculate the density at the
ICB, the melting temperature using Equation \ref{eqn.lindeman}, and
$T_c$ using Equation \ref{eqn.tempprofile}.\\

The temperature profile of the inner core will depend on whether it,
too, is convecting \citep{Buffett09}.  Such convection may be
sufficiently torpid so as not to contribute to the dynamo, but as long
as overturn is rapid compared to the timescale of inner core growth
($10^9-10^{10}$~yr), the temperature profile will be an adiabat.  This
assumption maximizes the transport of sensible heat into the outer
core, and hence minimizes the rate of inner core growth.

For a fully convecting core, the heat conducted along the the adiabat
at the CMB is (neglecting compressibility effects)
\begin{equation}
Q_K \approx \frac{4 \pi R_c^2 k \alpha_c g_c T_c}{c_p},
\end{equation}
where $k$ is the thermal conductivity and $g$ is the local
gravitational acceleration.  If $Q_c$, the heat flow from the core to
the mantle across the CMB, is less than this, the temperature profile
will be sub-adiabatic to the degree required to equalize the heat
flows.  Whether this layer is completely stratified will depend on
non-thermal (i.e., light element) buoyancy forces.  In the absence of
a growing inner core, however, such a layer will be thermally
stratified.  Geomagnetic variation limits the thickness of any upper
stratified layer in the Earth's core to $<100$~km \citep{Gubbins07},
if one exists at all \citep{Stanley08}.\\

If the stratified layer is thin and makes a negligible contribution to
the sensible heat flow, the heat flow through it is constant and its
temperature profile is simple. The location $\Re_*$ and temperature
$T_*$ of the boundary between the convective and conducting regions is
found by simultaneously solving for the heat flow and temperature:
\begin{equation}
\label{eqn.stratlayer}
\Re_*^3 = \left(\frac{D_c}{R_c}\right)^3\frac{Q_c(T_c)}{8 \pi k D_c T_*}
\end{equation}
and
\begin{equation}
\label{eqn.ttop}
T_c = T_* - \frac{Q_c (T_c)}{4\pi k R_c}\left(\frac{1}{\Re_*}-1\right),
\end{equation}
and using Equation \ref{eqn.tempprofile}, i.e.
\begin{equation}
T_* = T_i \exp \left[\left(\Re^2-\Re_*^2\right)\left(\frac{R_c}{D_c}\right)^2\right],
\end{equation}
where the dependence of $Q_c$ on $T_c$ must be accounted for (\S
\ref{sec.heat}), and we evaluate the thermal scale length at the CMB.
In the event that stratification does occur, we compute the entropy
terms for the {\it convecting part of the core only}.  This is done by
re-scaling the radius and mass of the core in the entropy equations to
the size and mass of the convecting zone.  $T_c$ becomes the
temperature $T_*$ at the top of the convecting zone, and we assume
that since the adiabat is self-similar, such that $dT_*/T_* \approx
dT_c/T_c$ (effectively ignoring small differences due to changes in
the thickness of the convective zone itself).\\

\subsection{Equation of entropy production}

The condition for maintenance of convective motions for a dynamo can
expressed as a balance between sources and sinks of entropy in the
core, e.g \citet{Labrosse07} and \citet{Nimmo09}:
\begin{equation}
\label{eqn.entropy}
\Phi + E_{K} = E_R + E_S + E_G + E_L,
\end{equation}
where $E_K$ is the entropy sink due to conduction along the adiabat,
$E_R$ the production of entropy by internal heating; $E_S$ that due to
core cooling (loss of sensible heat); $E_G$ the entropy production
from buoyancy generation; and $E_L$ the entropy production due to
latent heat generated by crystallization of the core.  The decay of
radioisotopes is a possible internal heat source.  $^{60}$Fe will
decay completely ($t_{1/2}=2.6$~Myr) \citep{Rugel09} before planets
finish accreting, but $^{40}$K ($t_{1/2} = 1.25$~Gyr) may be a heat
source in Earth's core \citep{Nimmo04}.  Values of the partition
coefficient of K between silicate and Fe/Ni alloy melts have been
discrepant between different high pressure experiments, possibly for
technical reasons \citep{Li03}, but new results limit the abundance of
K to a few tens of ppm, with no evidence for an increase in solubility
with pressure \citep{Corgne07}. In the absence of radiogenic heating,
\begin{equation}
\label{eqn.phi}
\Phi = E_L + E_S + E_G  - E_K.
\end{equation}
We calculate $\Phi$, divide by the outer core mass to obtain $\phi$,
and use Equations \ref{eqn.bfield2}, \ref{eqn.bsurf}, and
\ref{eqn.rem2} to estimate the magnetic field intensity and magnetic
Reynolds number.

Expressions for entropy production (dissipation) have been derived and
discussed extensively in the literature, e.g. \citet{Lister03}.
Following \citet{Labrosse07} and \citet{Nimmo09} we derive expansions
in terms of the size of the inner core.  The term from the loss of
sensible heat comes form the thermodynamic identity $dS = c_p dT/T$
under constant pressure and no heats of reaction.  Thus
\begin{equation}
E_S = -\int \left(\frac{1}{T_c} - \frac{1}{T}\right) \rho c_p \frac{\partial T}{\partial t} dV,
\end{equation} 
where the integration is over the volume of the entire core.  In the
absence of an inner core, and to fifth order in $R_c/D$ and $R_c/L$,
\begin{equation}
\label{eqn.es_nocore} 
E_S \approx - \frac{2}{5}c_p M_c  \left(\frac{R_c}{D}\right)^2 \left[1 + \frac{2}{7}\left(\frac{R_c}{D}\right)^2 + \frac{6}{35}\left(\frac{R_c}{L}\right)^2\right]\frac{dT_c}{dt},
\end{equation}
where $M_c$ is the total mass of the core.  We assume that subsolidus
convection continues the adiabatic temperature profile into the inner
core \citep{Buffett09} and ignore the effect of small changes in the
temperature and pressure length scales due to differences in
compressibility and density between the liquid and solid ($\epsilon$)
phases.  This assumption maximizes the contribution of the inner core
to the entropy and energy terms.  In the case of a mostly solid core,
a detailed treatment of the heat flow from the inner core would be
warranted.  However, we shall show that in planets much larger than
the Earth, inner cores remain relatively small, or do not grow at
all.\\

The entropic term for the latent heat due to core growth is;
\begin{equation}
E_L = 4\pi R_i^2 \rho_i \Delta S \left(\frac{T_i}{T_c} - 1\right)\frac{dR_i}{dt},
\end{equation}
where $\Delta S$ is the specific entropy of fusion.  Assuming that the
core grows outwards,
\begin{equation}
\frac{dR_i}{dt} = -\frac{D^2}{2R_i\left(\Delta -1\right)}\frac{1}{T_c}\frac{dT_c}{dt},
\end{equation}
where $\Delta$ is the ratio of the temperature-pressure slopes of the
solidus to the adiabat, evaluated at $R_i$.  Then,
\begin{equation}
E_L = -\frac{3}{2}\frac{M_c \Delta S }{\Delta -
1}\frac{\rho_i}{\bar{\rho}} \Re \left(1 - \Re^2 \right) \left[1 +
\left(\frac{R_c}{D}\right)^2 \frac{1-\Re^2}{2}\right]\frac{1}{T_c}\frac{dT_c}{dt},
\end{equation}
where the ratio of the density at the top of the inner core to the
mean density is obtained directly from the interior model (\S
\ref{sec.interior}).\\

The entropic term for the release of buoyant fluid (density deficit
$\Delta \rho$) during the crystallization of the core is
\citep{Labrosse07};
\begin{equation}
E_G \approx -\frac{3 \pi G M_c \Delta \rho D^2}{\Delta - 1}\frac{\rho_i}{\bar{\rho}}\Re \frac{F}{F'}{T_c^2}\frac{dT_c}{dt},
\end{equation}
where $\Delta \rho$ is the density change in the fluid due to the
presence of light elements, and $F$ and $F'$ are the dimensionless functions
\begin{eqnarray*}
F = \frac{1}{5} - \frac{1}{3}\Re^2 + \frac{2}{15}\Re^{5} + \\
\left(\frac{R_c}{L}\right)^2\left[\frac{2(1-\Re^7)}{21} - \frac{1-\Re^4}{6}\right] + \\
\left(\frac{R_c}{L}\right)^4\left[\frac{1-\Re^9}{27} - \frac{1-\Re^6}{18}\right],
\end{eqnarray*}
\begin{equation}
F' = 1 - \Re^3 - \frac{3}{5}\left(\frac{R_c}{L}\right)^2\left(1-\Re^5\right) - \frac{3}{14}\left(\frac{R_c}{L}\right)^4\left(1-\Re^7\right),
\end{equation}
c.f \citet{Nimmo09}.  To derive this, we ignore work done by expansion
of the core \citep{Lister03}, and assume that the energy is
dissipated at temperature $T_c$.\\

The entropy production by the conduction of heat along the adiabat is 
\begin{equation}
E_K = \int_{0}^{R_c} 4\pi r^2 k \left(\frac{1}{T}\frac{\partial T}{\partial r}\right)^2 dr
\end{equation}
\citep{Lister03}.  This is evaluated to be
\begin{equation}
E_K = \frac{12 k M_c}{5\bar{\rho}D_c^2}\left(\frac{R_c}{D_c}\right)^2\left(1-\Re^5\right).
\end{equation}
To more accurately reflect the temperature profile, to which
conduction is sensitive, we have evaluated the temperature scale
length at the CMB, rather than the core center.  The other terms in
the entropy equation depend on $D$ to a lesser degree and are
evaluated over the entire core, not just the CMB.\\

\subsection{Core cooling rate \label{sec.heat}}

We solve for the core cooling rate $dT_c/dt$ by balancing the heat flow:
\begin{equation}
\label{eqn.qbalance2}
Q_S + Q_L + Q_G = Q_c,
\end{equation}
where the terms correspond to sensible heat, latent heat,
gravitational energy, and heat transported through the CMB,
respectively.  The left-hand terms are all proportional to $dT_c/dt$.
The sensible heat term is
\begin{equation}
Q_S = \int \rho c_p \frac{\partial T}{\partial t} dV.
\end{equation} 
Assuming the adiabat continues into the inner core, this is 
\begin{equation}
\label{eqn.sensibleheat}
Q_S \approx c_p M_c \left[1 + \frac{2}{5}\left(\frac{R_c}{D}\right)^2 + \frac{4}{35}\left(\frac{R_c}{D}\right)^4 + \frac{12}{175}\left(\frac{R_c^2}{DL}\right)^2\right]\frac{dT_c}{dt}.
\end{equation}
The latent heat term is
\begin{equation}
\label{eqn.latentheat}
Q_L = \frac{3}{2}\frac{M_c \Delta S \Re}{\Delta - 1} \left(\frac{D}{R_c}\right)^2 \exp \left[(R_c/D)^2(1-\Re^2)\right]\frac{dT_c}{dt},
\end{equation}
while the gravitational term is,
\begin{equation}
\label{eqn.gravheat}
Q_G \approx -\frac{3 \pi G M_c \Delta \rho D^2}{\Delta - 1}\frac{\rho_i}{\bar{\rho}}\frac{F}{T_c}\frac{dT_c}{dt},
\end{equation}
We compare $Q_c$ to the heat $Q_K$ that can be carried by conduction
along the core adiabat:
\begin{equation}
\label{eqn.heatconduction}
Q_K = \frac{8\pi R_c^3 k T_c}{D_c^2}.
\end{equation}
If $Q_c > Q_K$ then the entire core has thermally driven convection.
If $Q_c < Q_K$ then cooling is insufficient to maintain convection in
the entire core and a stratified, conducting layer develops (\S
\ref{sec.tempprof}).  In our calculation of the rate of sensible heat
loss, we have neglected the small corrections due to departures of the
temperature profile from the purely adiabatic case.\\

\subsection{Heat transport in the mantle \label{sec.mantle}}

We assume that the mantle boundary layer at the CMB develops
independently of the flow in the rest of the mantle and its thickness
is controlled by a local Rayleigh number Ra, as opposed to the
whole-mantle Ra \citep{Nimmo00}.  Heat is removed from the core
according to;
\begin{equation}
\label{eqn.qcmb}
Q_c = 4\pi R_c k_m {\rm Nu}_c \left(T_c - T_l\right),
\end{equation}
where $T_l$ is the temperature of the lower mantle and $k_m$ is the
thermal conductivity of the mantle.  The Nusselt number is
\begin{equation}
\label{eqn.nussult}
{\rm Nu}_c \approx ({\rm Ra}_c/{\rm Ra}_*)^{\delta},
\end{equation}
where $\delta \approx 0.3$ \citep{Schubert01}.  For a fluid heated
from below,
\begin{equation}
\label{eqn.racmb}
{\rm Ra}_{c} = \frac{\rho g \alpha \left(T_c - T_l\right) d^3}{\kappa \eta_c}
\end{equation}
where the gravitational acceleration $g$ and viscosity $\eta$ are
evaluated at the CMB, and $d = R_p - R_c$.  In the case of a partially
stratified layer, we substitute $T_t$ for $T_c$, where $T_t$ is the
temperature at the top of the stratified, conducting zone.

The heat transport by subsolidus convection in the mantle is;
\begin{equation}
\label{eqn.mantleheat}
Q_{m} = \frac{4 \pi R_p^2 {\rm Nu}_m k \left(T_m - T_p\right)}{R_p - R_c},
\end{equation}
where $T_p$ is the surface temperature.  For a uniform internal heat
source,
\begin{equation}
\label{eqn.raman}
{\rm Ra_m} = \frac{\rho g H  \alpha d^5}{k \kappa \eta_m},
\end{equation}
where $H$ is the total specific heat generation (per unit mass).
$\eta_m$ is evaluated at the mantle reference temperature, taken to be
halfway along the adiabat.  We make the approximation that heat from
the CMB is mixed rapidly and uniformly into the mantle (e.g., by
plumes) and thus $H = H_c + H_r$, where $H_c = Q_c/M_m$, where $M_m$
is the mass of the mantle, and the radiogenic heat production $H_r$ is
given by;
\begin{equation}
H_r = \Sigma_i C_i h_i e^{-t/\bar{t}_i},
\end{equation}
$C_i$ is the initial concentration of the $i$th radioactive isotope,
$h_i$ is the specific heat production, and $\bar{t}_i$ is the mean
life.  The concentrations adopted for the long-lived radioisotopes
$^{40}$K, $^{232}$Th, $^{235}$U, and $^{238}$U are for the case of an
``undepleted'' terrestrial mantle \citep{Ringwood91} given in Table 2
of \citet{Kite09}.  The contribution due to mantle cooling is;
\begin{equation}
Q_T = -c_m M_m \frac{dT_m}{dt},
\end{equation}
where $c_m$ is the is the specific heat capacity of the mantle.

We adopt the viscosity law for a stress exponent $n = 1$ material,
appropriate for the plate-tectonic regime \citep{Nimmo00}:
\begin{equation}
\eta_m = \eta_* \exp \left(b \tau_m /T\right),
\end{equation}
where $b \approx 17$ \citep{Karato01}, $\tau_m$ is the peridotite
solidus, and $\eta_*$ is a reference viscosity.  In Equation
\ref{eqn.racmb}, we evaluate $\eta$ at the ``film'' temperature, which
is intermediate that of the lower mantle and top of the core
\citep{Manga01}.\\

Convection in the silicate mantle may involve a mobile lid (i.e.,
plate tectonics) or a stagnant lid, and in our Solar System we have
Earth-mass examples of each (Earth and Venus).  Whether one mode or
another will be more likely on larger planets will depend on the
relative magnitudes of lithosphere stresses \citep{Valencia07}, the
availablity of water to weaken faults \citep{Nimmo98}, surface
temperatures \citep{Lenardic08}, and the formation of buoyant,
unsubductable crust \citep{Kite09}.  In our model, the mode of mantle
convection controls (a) the temperature boundary condition at the
mantle side of the CMB; and (b) the Nussult-Rayleigh number
relationship in the mantle.\\

In the case of a mobile lid, we assume that subducted slabs reach the
lower mantle, have a temperature equal to the mean mantle temperature
$T_m$, and directly cool the core.  Thus for Equation \ref{eqn.qcmb}
we set $T_l = T_m$.  We the use the same Nu-Ra relation as in Equation
\ref{eqn.nussult}.  In the stagnant lid case, the lower mantle
temperature is set to
\begin{equation}
T_l = \theta \exp \left[\int_{R_c}^{R_p} \frac{\alpha g}{c_p} dr\right],
\end{equation}
where the $\theta$ is the potential temperature of the mantle, and the
relationship between mean and potential temperatures is taken to be
\begin{equation}
\theta = T_m exp \left[-\frac{1}{2}\int_{R_c}^{R_p} \frac{\alpha g}{c_p} dr\right].
\end{equation}
We adopt the relationship between Nusselt number, Rayleigh number, and
viscosity contrast $N$ across the conducting lid (Frank-Kamenetskii
parameter) given by \citet{Solomatov00};
\begin{equation}
{\rm Nu} = a N^{-(1+\beta)}{\rm Ra}^\beta.
\end{equation}
where $a \sim 1$, and boundary layer stability analysis shows that for
a material with stress exponent $n$, $\beta = n/(n+2)$
\citep{Solomatov95}.  We use $\beta = 0.6$.\\

\section{Results}

\subsection{Parameter values  and a nominal Earth-mass case \label{sec.sensitivity}}

Our numerical experiments revealed that the predicted history of the
dynamo in a planet with an Earth-like interior is sensitive to: (1)
mantle viscosity; (2) the thermal expansivity of iron (related to its
compressibility); (3) the thermal conductivity of Fe; (4) initial core
temperature; and (5) the heat flow across the CMB.  The first is
constrained by the requirement to reproduce the size of and pressures
in the Earth's core (\S \ref{sec.interior}).  We parameterize the last
by using different values of the critical Rayleigh number Ra$_*$.
Althhough Ra$_*$ is of order $10^3$ \citep{Nimmo00,Turcotte02}, we
consider a range of values that encapsulate the complex uncertainties
of the thermochemical boundary layer at the CMB.\\

{\it Mantle viscosity:} Our viscosity law requires a solidus and a
reference viscosity.  We fit a Lindemann melting law to the solidus
data of \citet{Zerr98} for the lower mantle, using the Mg-perovskite
equation of state to relate $P$ to $\rho$: We found a best-fit
Gr\"{u}neisen parameter of 1.45.  The reference viscosity of a
hydrated Earth-like mantle was adjusted to achieve a present potential
temperature $\theta$ of 1750~K.  This is the present temperature
beneath mid-ocean ridge basalts (1623~K) \citep{Herzberg10} to which
has been added 130~K associated with the olivine-spinel transition in
the deep mantle (latent heat of reaction = 160 kJ~kg$^{-1}$)
\citep{Ito71}.  The derived reference viscosity, $\eta_* = 6 \times
10^{19}$~Pa~s, is somewhat above inferred values for the terrestrial
asthenosphere where partial melting may occur
\citep{Solomatov00,Dixon04,Fjeldskaar07,Bills07,James09}.  The
viscosity of a dehydrated mantle is expected to be significantly
higher. The effect of water on viscosity depends on the dominant creep
mechanism, and the partitioning of water amongst the mineral phases
that are present.  Hydration of olivine in the upper mantle could
lower the viscosity by a factor of $\sim$500 \citep{Hirth96}, but the
degree of hydration, and concomitent decrease in viscosity, could be
much less in the lower perovskite mantle \citep{Bolfan00}.  We
selected a value based on the requirement that a simulation of Venus
(0.815 \mearth~and T$_p$ = 785~K) predicts no dynamo at the present
time.  [The slow rotation of Venus cannot be solely responsible for
the absent dynamo \citep{Zhang95}.]  If Venus has had a stagnant lid
over its entire history, then the required viscosity enhancement for
the Venusian dynamo to cease by 4.5~Gyr is a factor of 6.  If its lid
was primarily mobile, then the required viscosity enhanceent is a
factor of 200.  We conservatively adopt a factor of 10, recognizing
that higher values are possible.\\

{\it Fe thermal conductivity:} Thermal conduction competes with
advection in removing heat from the core.  Higher thermal conductivity
will suppress convection, and thus the dynamo, and will stratify some
or all of the liquid core.  The thermal conductivity of iron is
expected to depend on pressure, temperature, and the presence of light
elements, but not on temperature \citep{Berman76}.  The value often
used in core models is 50-60~W~m$^{-1}$~K \citep{Labrosse03}.  Thermal
conductivity can be determined from measurements of electrical
conductivity using the Wiedemann-Franz-Lorenz (WFL) law.  Assuming the
WFL law applies at high pressure, electrical conductivity measurement
at 140~GPa \citep{Keeler69} and 200~GPa \citep{Bi02} imply $k
\sim$100~W~m$^{-1}$~K$^{-1}$, c.f., \citet{Anderson98}.  In contrast,
\citet{Stacey07} argue for a value of 28-29~W~m$^{-1}~$K$^{-1}$ based
on the expectation that scattering of electrons by 3d orbitals (which
contributes to resistivity) will increase faster with pressure than
the densities of 3d and 4s states (which contribute to conductance).
The electrical conductivity measurements of \citet{Matassov77} show
that light elements can significantly reduce thermal conductivity
\citep{Anderson98}, e.g., a factor of two reduction for Fe$_3$Si
\citep{Manga96}.  A low value of $k$ is an alternative mechanism to
radiogenic heat to avoid core stratification \citep{Stacey07}.\\

{\it Initial core temperature:} We presume that the cores of more
massive planets will be initially hotter because of the conversion of
a greater amount of gravitational potential energy into heat.  A
hypothetical planet with a small core that grows concurrently by the
descent of iron from a cold surface will have a final temperature $T_0
\approx (3G M_p)(10 R_p c_p) \sim 22100 (M/M_{\oplus})^{2/3}$~K.
These core temperatures are unrealistic because much of the
gravitational potential energy will be dissipated in and carried back
to the surface by a partially molten silicate mantle.  Instead, we set
the initial temperature at the top of the core relative to the melting
temperature of Mg-perovskite (MgSiO$_3$), effectively the liquidus
temperature of the mantle, at the CMB.  Core temperatures reaching
this would result in the formation of a massive magma ocean whose low
viscosity would quickly cool a hotter core.  Alternatively, our
initial state can be considered the epoch when any such magma ocean
solidifies.  The experimental data and {\it ab initio} calculations of
the MgSiO$_3$ melting temperature \citep{Stixrude09} show that $\tau$
is well described to 136~GPa by the relationship $\tau \approx$
5400(P/136 Gpa)$^{0.367}$.  This is equivalent to a Lindeman law and
$\gamma=1.44$, c.f. \citet{Akaogi93}, and we simply extrapolate this
to higher pressures. The mantle is given its steady-state temperature
as its initial temperature.\\

\begin{figure}
\noindent\includegraphics[width=2.5in]{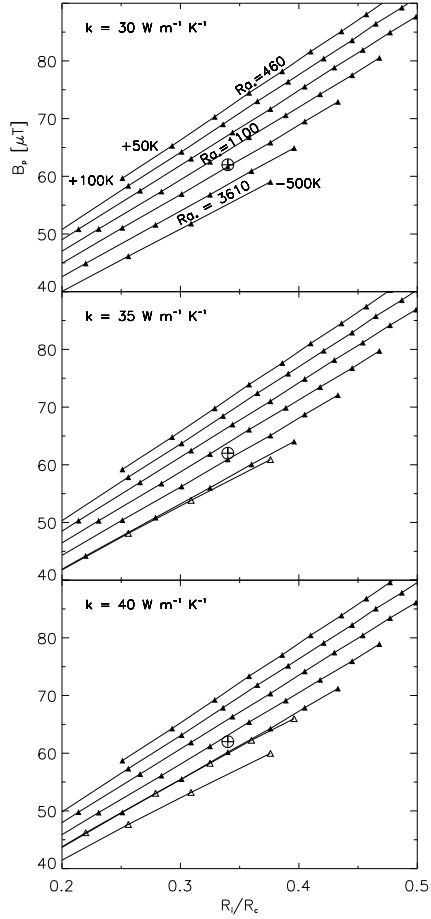}
\caption{Predicted inner core radius and surface field for
4560~Myr-old Earth-mass planets with different values of initial core
temperatures $T_c$, critical Rayleigh number Ra$_*$ (which
parameterizes the uncertain efficiency of heat transfer across the
CMB), and thermal conductivity of Fe $k$. Each curve represents a
series of model runs with a given Ra$_*$ of (top to bottom) 460, 600,
800, 1100, 1560, 2300, and 3610.  Each point along a curve is
represents a run with a different value of $T_c$.  The far right point
represents a value 500~K below the lower mantle liquidus (5400~K).
Each point further to the left represents an increase of $T_c$ in 50~K
steps until 100~K above 5400~K.  Solid points represent runs where the
entire core is convecting at 4560~Myr.  Open points represent
partially stratified cores.  All runs assume plate tectonics and a
surface temperature of 288~K.  The present inner core radius and mean
surface field of the Earth is plotted as the large circle in each
pane.  The mean was calculated using a dipole moment of $8 \times
10^{22}$ A~m$^2$.  We assume that the magnetic field is a pure dipole,
and thus our model overpredicts the mean field strength at the surface
of the Earth. }
\label{fig.sensitivity}
\end{figure}

Figure \ref{fig.sensitivity} plots the predicted inner core size and
the surface magnetic field of the Earth at 4560~Myr for different
values of the initial core temperature (from 100~K above to 500~K
below the MgSiO$_3$ melting temperature), critical Rayleigh numbers
between 460 and 3610, and three different values of the thermal
conductivity (30, 35, and 40 W~m$^{-1}$K~$^{-1}$).  We calculate a
present surface-average terrestrial field strength using a dipole
moment of $8 \times 10^{22}$~A~m$^2$.  To compare with the partially
non-dipolar field of the Earth, we multiply our model predictions by
80\% \citep{Lowes07}.  Because 80\% is the measured dipole fraction at
the surface, and the fraction at the core will be smaller, our
predictions should over-estimate the strength of the Earth's current
field.  At $k > 45$~W~m$^{-1}$~K$^{-1}$ the core becomes partly
stratified, inconsistent with observations of magnetic field
variability \citep{Gubbins07,Stanley08}.  At lower values of $k$, the
present core size and magnetic field strength are can be reproduced by
combinations of a range of critical CMB Rayleigh numbers and initial
core temperatures.  We adopt $k = 35$~m$^{-1}$~K~$^{-1}$, $Ra_* =
1100$, and an initial core temperature of 275~K below the MgSiO$_3$
melting temperature at CMB pressure.  We use these parameter values
for all subsequent calculations. \\

\begin{figure}
\noindent\includegraphics[width=2.5in]{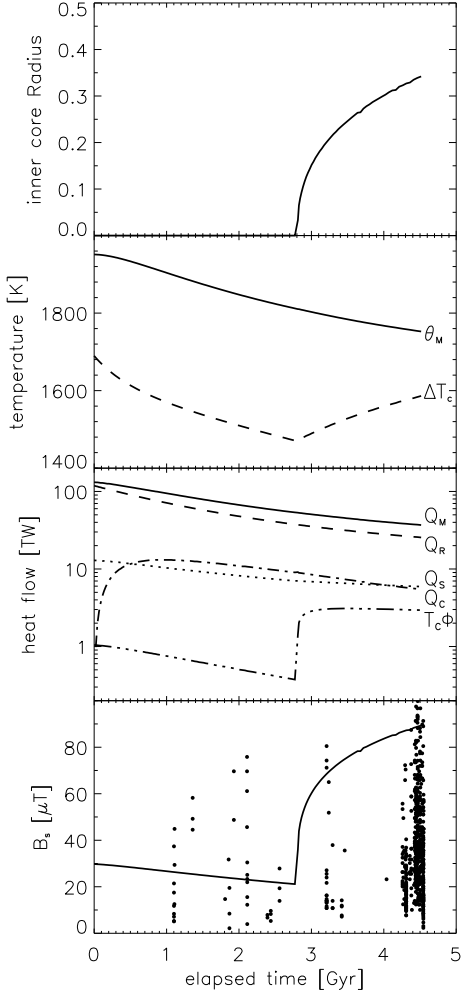}
\caption{Predicted evolution of the Earth and its dynamo.  Top to
bottom: inner core radius relative to outer core; mantle potential
temperature $\theta$ (solid) and temperature contrast across the CMB
$\Delta T_c$ (dashed); total surface $Q_m$ (solid), radiogenic $Q_r$
(dashed), mantle cooling $Q_s$ (dot-dash), and core $Q_c$ (dotted)
heat flows, and $\phi T_c$, the power available for the dynamo
(dash-triple-dot); average surface magnetic field.  In the last pane,
606 measurements from the IAGA paleointensity database
\protect{\citep{Biggin09}} and three recent measurements in 3.45 Ga rocks
\protect{\citep{Herrero09,Tarduno10}} are plotted.  Only paleointensity
measurements older than 10~Myr, having an error less than 50\%, and
based on Thellier or Shaw-Thellier techniques with a pRTM check, are
shown.
\label{fig.evolution}}
\end{figure}

Figure \ref{fig.evolution} shows the calculated 4.56~Gyr evolution of
a 1 \mearth~planet.  The mantle potential temperature decreases by
200~K, consistent with paleothermometry based on the chemistry of
Archean and Proterozoic non-arc basalts \citep{Herzberg10}.  The
temperature contrast across the CMB decreases until the onset of inner
core formation, then rises slightly, as the core is heated by latent
heat from the inner core, to a present value of around 1600~K, only
slightly higher than current estimates \citep{Tateno09}.  Calculated
mantle heat flow, radiogenic heat flow, and core heat flow are all
consistent with current estimates \citep{Buffett02,Lay08}, but the
Urey number at 4.56~Gyr is $\sim$0.8, significantly higher than in
some models \citep{Korenaga08}.  The predicted evolution of the
surface field (or the equivalent dipole moment) is comparable to that
of other models \citep{Labrosse07,Aubert09,Breuer09}.  It also does
not conflict with paleomagnetic data, represented by a filtered sample
of the IAGA paleointensity database \citep{Biggin09} as well as three
recent measurements in 3.45~Ga rocks \citep{Herrero09,Tarduno10}.\\

\subsection{Inner cores versus iron snow \label{sec.snow}}

\begin{figure}
\noindent\includegraphics[width=3in]{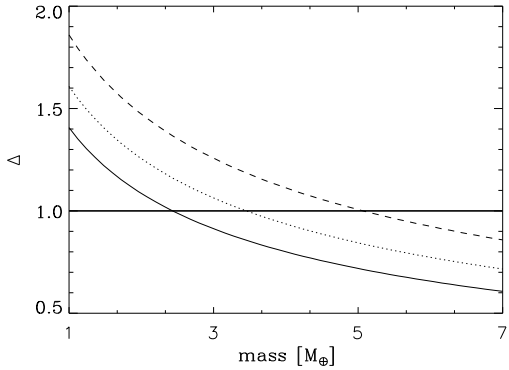}
\caption{Predicted ratio $\Delta$ of the temperature-pressure slopes
of the adiabat versus the solidus in Earth-like planets with different
total masses (Earth units).  For $\Delta > 1$, a solid inner core
forms.  When $\Delta < 1$, iron condenses at the top of the core.  In
the latter case the release of latent heat stratifies the core and a
dynamo is not expected.  Solid line: no inner core; dotted line: 50\%
inner core by mass; dashed line: nearly completely solid core.}
\label{fig.delta}
\end{figure}

Whether a core grows outward from the center of the planet, or inwards
from the top of the core depends on $\Delta$, the ratio of the
pressure-temperature slopes of the adiabat and solidus.  This value
will vary with depth in the core, but is prudently evaluated at the
inner core boundary or, when no core is present, at the center of the
planet.  For an adiabatic temperature profile,
\begin{equation}
\label{eqn.delta}
\Delta = \frac{2\left(3\gamma - 1\right)K_1 \left(1 + \ln (\rho_i/\rho_0)\right)}{\gamma \left[dP/d(\rho/\rho_1)\right]}.
\end{equation}
The zero-pressure compressibility $K_1$ appears as a linear factor in
the EOS, therefore $\Delta$ ultimately depends {\it only} on the
Gr\"{u}neisen parameter $\gamma$, the pressure-derivative terms in the
EOS, and the pressure at the inner core boundary.  Because gravity is
comparatively low in the core, the pressure is relatively insensitive
to the size of the core, and the total mass of the planet is the most
important parameter.  $\Delta$ will decrease with mass due to the
increasing incompressibility of Fe (including the TFD effect) at high
pressures.  $\Delta < 1$ for planets larger than
2.5-3~\mearth~(Figure \ref{fig.delta}).\\

\begin{figure}
\noindent\includegraphics[width=3in]{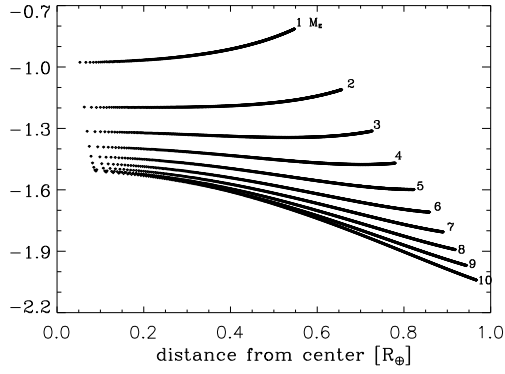}
\caption{Logarithmic ratio of temperature to solidus with radius in
entirely liquid (Fe) cores in planets of 1 (top) to 10 (bottom) Earth
masses.  Unimportant constants have been ignored.  Fe solidification
will occur first at the minimum of each curve.  Planets with masses
less than about 2.5~\mearth~will form solid inner cores.  In more
massive planets, we predict that iron ``snow'' will condense near or
at the top of the core.  The concomitant release of latent heat and
the absence of buoyancy forces will stratify the core.}
\label{fig.ratio}
\end{figure}

Figure \ref{fig.ratio} shows the logarithmic ratio of the temperature
to the melting point versus depth in the cores of
1-10~\mearth~planets.  An unimportant constant has been ignored in
each case.  The minima in these curves is where iron will first
solidify.  Above 2~\mearth~the minimum shifts from the center
(corresponding to inner core growth) to a position near or at the CMB.
In these cases, iron ``snow'' will freeze out at or near the top of
the core.  Except under an exceptionally narrow range of conditions
when the temperature coincides with the solidus throughout much of the
core (Figure \ref{fig.ratio}), iron crystals or slabs will first form
at or near the top of the core and sink, and the resulting release of
light elements and transport of latent heat will stabilize the liquid
core and shut down convection \citep{Williams09}.  This effect has
been discussed for several small bodies in the Solar System
\citep{Hauck06,Stewart07,Chen08,Bland08}.

Because there is no {\it net} release of latent heat, the core will
continue to cool and iron will crystallize over an expanding depth
range (Figure \ref{fig.ratio}).  Double diffusive convection can drive
a ``fingering'' instability in a stratified fluid where two gradients,
e.g. a thermal gradient and compositional gradient, oppose each other,
but are associated with substantially different diffusivities
\citep{Huppert81,Huppert84}.  Double diffusive convection has been
included in a recent model of Mercury's dynamo \citep{Manglik10}.  In
our case, the release and absorption of latent heat by crystallizing,
sinking, and melting grains of iron will create a slightly
subadiabatic temperature gradient, and buoyancy forces due to
temperature will not play a role.  Double diffusive convection will
only occur if (a) the diffusion of the light elements is faster than
the iron crystals, and (b) the crystals are well-coupled to the fluid,
i.e. there is a high density of nucleation sites and crystals are
small.  Even if convection ensues, it will do so in discrete layers
each of which has a Rayleigh number of order the critical value
\citep{Huppert84}.  Boundary layer theory for finite-amplitude
convection \citep{Turcotte02} shows that the magnetic Reynolds number
of each layer will be of order $(\kappa/\lambda)$Ra$^{2/3}$, or $\sim
10^{-3}$, well below that required to sustain a dynamo.  If the
crystals are not coupled to the fluid, no convective instability will
occur. and instead crystals will steadily settle through the fluid.  It
is not clear that a dynamo occurs under these conditions, either.

\subsection{Variation in the dynamo history of Earths and super-Earths}

\begin{figure}
\noindent\includegraphics[width=3in]{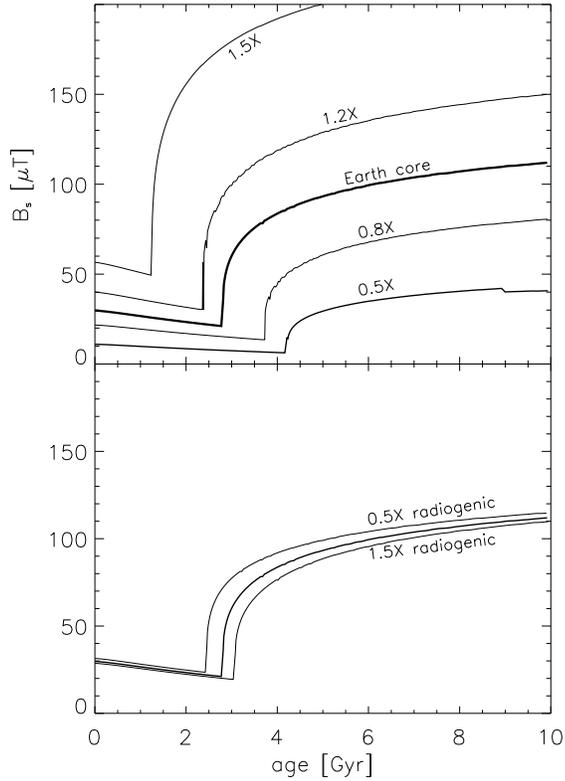}
\caption{Predicted history of the average surface magnetic field of
Earth-mass planets with plate tectonics and surface temperatures of
288~K.  Dynamos are assumed to produce pure dipole fields.  (a) Top:
planets with core masses between 0.5 and 1.5$\times$ that of Earth.
(b) Bottom: Planets with an Earth-size core but initial radiogenic
element abundances that are 0.5 or 1.5 $\times$ the terrestrial case
(heavy solid line).}
\label{fig.sizeandrad}
\end{figure}

For a given total mass, the mass of a planet's core may differ as a
result of variation in the ratio of Fe to Si in the primordial disk
\citep{Neves09}, the efficiency of core formation \citep{Elkins08} or
impact stripping of the mantle \citep{Benz88,Benz07}.  The ability of
impacts to remove substantial amounts of the mantle from Earth-mass
and larger planets might be limited, and thus massive Mercurys may
not be plausible \citep{Marcus10}.  Figure \ref{fig.sizeandrad}a shows
the predicted dynamo history of Earth-mass planets with core sizes
between 0.5 and 1.5 times that of the Earth.  The average magnetic
field is higher on planets with larger cores, mostly due to the
smaller $r^{-3}$ attenuation of the dipole field outside the core.

The initial abundance of heat-producing radioactive nuclides
($^{40}$K, $^{232}$Th, $^{235}$U, and $^{238}$U) may also vary between
planets and between systems as a result of chemical and thermal
processing of solids in planet-forming disks, and galactic chemical
evolution \citep{Kite09}.  K is a volatile element and the abundance
of $^{40}$K will vary with the location of a planet's accretion zone,
with those planets (e.g. Mars) forming further out having a higher
abundance \citep{Lodders97}.  The Galactic abundances of these
isotopes relative to major planet-forming elements such as Si are
predicted to have decreased substantially with time, and planets of a
given age may accrete with substantially different inventories, but
the difference between planets of different ages observed {\it at the
present time} is much smaller \citep{Kite09}.  Figure
\ref{fig.sizeandrad}b shows the predicted dynamo evolution for planets
with 50\% and 150\% of the Earth's estimated initial inventory of
radionuclides.  The predicted surface field is slightly lower with a
higher abundance of radionuclides, but is relatively insensitive to
this parameter

\begin{figure}
\noindent\includegraphics[width=3in]{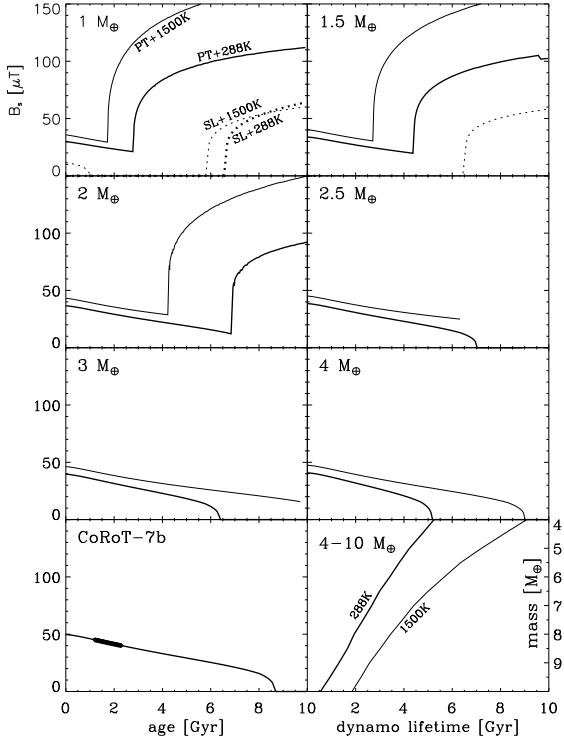}
\caption{Predicted history of the average surface magnetic field of
planets with masses of 1-10~\mearth~and the $\sim$4.8~\mearth~planet
CoRoT-7b.  Absence of a line indicates that the dynamo is inoperative.
Dynamos are assumed to produce pure dipole fields. Heavy solid lines:
Earth-like planets with plate tectonics and surface temperature of
288~K. Light solid lines: ``hot'' Earth with plate tectonics and
surface temperature of 1500~K.  Light dashed lines: Venus-like planets
with stagnant lid, $10 \times$ elevated mantle viscosity, and surface
temperature of 1500~K.  Heavy dashed lines: ``balmy'' Venus with a
surface temperature of 288~K.  For the case of CoRoT-7b, the surface
temperature is set to 1810~K, assuming efficient redistribution of
heat \protect{\citep{Leger09}}.  The thick part of the CoRoT-7 curve spans the
range of the system's estimated age (1.2-2.3 Gyr).  Lines terminate
when $\Delta < 1$ and ``iron snow'' forms at the top of the core, and
are absent when no dynamo is present or the simulation was
discontinued due to the predicted presence of a magma ocean.  The
behavior of planets more massive than 4 \mearth~is similar to the 4
\mearth~case.  The bottom right-hand panel is the predicted dynamo
lifetime as a function of planet mass $>$4~\mearth~and the two values
of the surface temperature.}
\label{fig.masstempmode}
\end{figure}

Figure \ref{fig.masstempmode} plots the predicted dynamo history of
1-10~M$_{\oplus}$ Earth-like planets with and without plate tectonics,
and surface temperatures of 288~K (Earth-like) and 1500~K (the solidus
of basalt).  Planets much larger than 1.5 \mearth~with stagnant lids
do not produce dynamos and even more massive planets have lower mantle
temperatures that exceed the estimated peridotite solidus.  Correct
modeling of planets with internal magma oceans is beyond the scope of
this study and we terminate those runs.  Only planets smaller than 2.5
\mearth~form inner cores ($< 2$ \mearth~for those with stagnant lids),
and the elapsed time before an inner core appears, corresponding to a
rapid increase in the strength of the magnetic field, increases with
planet mass.  The strength of the magnetic field at any given age
decreases with planet mass.  On Earth-mass planets with elevated
surface temperatures, the strength of the magnetic field is higher and
the interval before the onset of core formation is briefer.\\

Planets more massive than 2 \mearth~(like CoRoT-7b) have long-lived
dynamos produced by core cooling, but only if they have a mobile lid.
The duration of the dynamo decreases with planet mass and increases
with surface temperature.  The bottom right-hand panel of Figure
\ref{fig.masstempmode} extends these trends out to a mass of 10
\mearth.

\section{Discussion and Conclusions \label{sec.discussion}}

\subsection{Interpretation of trends and implications}

For an inner core to form, enough sensible heat must be removed for
the center temperature to reach the solidus.  We obtain an approximate
timescale $t_c$ for this by setting $Q_s = -Q_K$ and performing the
appropriate integration.  This gives
\begin{equation}
\label{eqn.growthtime}
t_c \approx \frac{M_c\Delta S}{8\pi kR_c} \left(1 -
\frac{D^2}{R_c^2}\ln \frac{\tau_{0}({\rm Fe})}{\tau_{c}({\rm
pv})}\right).
\end{equation}
where the solidii for Fe and perovskite are at central and CMB
conditions, respectively.  For $k=40$~W~m$^{-1}$~K$^{-1}$, and $R_c
\sim M_p^{1/4}$ (Figure \ref{fig.massradius}), the first factor is
$\approx 2 M_p^{3/4}$~Gyr.  The second factor is of order unity: Both
solidii increase with planet mass and so we expect $t_c$ to increase
slightly faster with planet mass than $M_p^{3/4}$.  This is born out
by our detailed calculations, i.e a 2 $M_{\oplus}$ planet takes more
than twice as long to form an inner core than an Earth-mass planet
(Figure \ref{fig.masstempmode}).

A maximum time $t_s$ for the complete solidification of the entire
core, after the inner core starts growing, is obtained by dividing the
total latent heat by the heat flow.  (In our cases, the heat flow
through the mantle tends to be close to or not much larger than the
conduction through the core itself.)  In this case, $t_s$ is dependent
on mass only through the pressure-dependences of certain material
properties of the core:
\begin{equation}
\label{eqn.freezetime}
t_s \approx \frac{c_p  \Delta S}{2 k \alpha G}.
\end{equation}
For standard values, $t_s \sim 50$~Gyr. The iron cores of rocky
planets with Earth-like composition will never completely solidify,
and solid inner cores never include more than about half of the total
core mass in 10~Gyr.  Thus, planets that do form inner cores ($\le$2
\mearth) will have dynamos that last longer than the main-sequence
life of the parent star.

In planets with masses of 2-3 \mearth, inner cores do not grow but
instead iron condenses as ``snow'' at the CMB, shutting down
convection and the dynamo within 10~Gyr.  In yet more massive planets,
the core never cools sufficiently for iron to condense.  These planets
have persistent dynamos produced by core cooling.  However, the
temperature contrast between the core and the mantle decays, and the
heat flow across the CMB eventually falls to the core conduction
value.  At that point, $\phi = 0$ and the dynamo ceases.\\

At a given planet mass, the surface field $B_p$ increases with
increasing core size, and a solid inner core formation forms earlier
(Figure \ref{fig.sizeandrad}a).  The former is a consequence of the
cubic distance attenuation law and a shallower CMB, as well as
increased heat flow across a thinner mantle.  The latter is a result
of more rapid cooling, but also higher pressure and a higher solidus
at the planet's center.  Dynamo history is much less sensitive to the
initial abundance of long-lived radionuclides in the planet's mantle
(Figure \ref{fig.sizeandrad}b) because of the self-regulating nature
of viscosity-dependent mantle convection \citep{Tozer72}.

$B_p$ generally decreases with planet mass.  Equations
\ref{eqn.bfield2} and \ref{eqn.bsurf} show why: The intensity of the
field in the core $B_c$ depends on the available entropy production
per unit mass $\phi$ and only weakly on core size.  All else being
equal, $\phi$ will decrease with mass because the core of a larger
planet has less surface area per unit mass and a thicker mantle
through which to lose heat.  If $\phi \sim R_c^{-1}$, this makes $B_c$
insensitive to planet mass.  The intensity of the field at the surface
will decrease as $R_c^{-3}$ and thus will roughly scale as
$M_p^{-3/4}$.\\

Higher surface temperature leads to elevated mantle temperatures and
lower viscosities.  Although the former lowers the temperature
contrast across the CMB, the latter more than compensates because of
the exponential decrease of viscosity with temperature.  The outcome,
increased magnetic field strength and dynamo lifetime, is clearly seen
in Figure \ref{fig.masstempmode}.\\

\subsection{CoRoT-7b}

Our model predicts that CoRoT-7b has a dynamo at its estimated age of
1.3-2.2~Gyr \citep{Leger09}, but only if it has a mobile lid.  In this
case, the dynamo is driven entirely by core cooling, and no inner core
is present.  In fact, it is the absence of iron solidification (as
``snow'') that allows a dynamo to persist.  Whether or not plate
tectonics operates on CoRoT-7b is debated
\citep{Valencia07,ONeill07,Kite09}, and the presence of a dynamo may
be one means of testing this.  However, a core dynamo is not the only
mechanism that can produce a magnetic field on such an object (see \S
\ref{sec.other}).\\

\subsection{Uncertainties and neglected effects \label{sec.uncertainties}}

{\it High-pressure material properties:} The properties of silicates
and Fe at pressures of hundreds or thousands of GPa are very
uncertain.  Although equations of state become dominated by electron
degeneracy pressure at pressures well above 1~TPa, other parameters,
namely $k$ and $\alpha$, are very important in determining whether a
dynamo operates.  \citet{Aitta06} has proposed that iron has a
tricritical point at around 800 GPa at which separate solid and liquid
phases disappear and the latent heat of fusion vanishes.  These
conditions are reached at the center of a 2.5~\mearth~planet and at
the CMB of a 6~\mearth~planet (Figure \ref{fig.massradius}).\\

{\it Mantle structure:} We assume a fully convecting mantle: A layered
mantle will support a higher temperature contrast for the same heat
flow \citep{McNamara00}, and this would tend to suppress dynamo
operation.  The lowest 150~km of the Earth's mantle (within the D''
layer) may be in the stability field of a post-perovskite (ppv) phase
\citep{Oganov04} and ppv may dominate in sufficiently cool
``super-Earth'' mantles \citep{Valencia09}.  If a magma ocean existed
at the base of the Hadean terrestrial mantle, release of latent heat
during crystallization would have suppressed the cooling of Earth's
core \citep{Labrosse07b}.  Our calculations indicate that the mantles
of planets with M $\ge$ 2.5\mearth~and stagnant lids are partially
molten.  We do not attempt to model the thermal evolution of such
objects but they are (at least initially) unlikely to support a
dynamo.\\

{\it Other effects of light element in the core:} Our model does not
include melting-point depression in the core by light elements,
particularly sulfur (S).  The effect of these is to delay inner core
formation relative to the pure Fe case.  In that sense our
calculations are optimistic in predicting dynamos.  (This is in
contrast to the case of small bodies where S can maintain a dynamo by
delaying the freezing of the entire core).  However, we choose our
initial core temperature based on the present size of the Earth's
inner core and thus, relative to Earth, this calibration may cancel
some of the effect of S.  We include the production of buoyancy flux
resulting from the exclusion of light elements from a growing solid
core.  However we assume that light elements are uniformly mixed into
the outer liquid core and do not include possible variations in light
element concentrations with planet mass.  The liquid silicate-liquid
metal partition coefficient of sulfur increases with pressure
\citep{Li03} but no experimental data is yet available for pressures
exceeding those at Earth's CMB.  If the trend continues to higher mass
(and for other elements), the power available for dynamos in the cores
of super-Earths may be smaller than our model predicts.\\

{\it Non-radiogenic sources of mantle heat:} Dissipation of tidal
strain could be an additional source of heat in the mantles of
non-synchronously rotating planets close to their parent stars
\citep{Barnes10}.  The timescale for rotational synchronization is a
few Myr \citep{Guillot96} while that of orbital circularization can be
Gyr \citep{Rasio96}, but both are strongly dependent on semimajor
axis.  Sufficiently intense heating during this period could lead to a
long-lived thermal anomaly in a mantle that influences the transport
of heat much like a stagnant lid.  We also ignore the potential
complication of heating by late giant impacts
\citep{Robert09,Arkani10}.\\

{\it Effect of the stellar magnetic field:} In the absence of an
ionosphere, the ambient magnetic field of the host star $B_*$ could
reach the planet's core: A sufficiently strong applied field will
inhibit convective motion in transverse directions.  This will occur
when the Stuart number
\begin{equation}
\label{eqn.stuart}
{\rm St} = \frac{\sigma B_*^2 R}{\rho v} \gg 1.
\end{equation}
Using the definition of Re$_m$ we re-write this as:
\begin{equation}
\frac{\sigma B_*^2 R^2}{\rho \lambda} \gg {\rm Re_m}.
\end{equation}
For typical values $\sigma \sim 5 \times 10^5$~S~m$^{-1}$, $\rho \sim
10^4$~kg~m, $\lambda \sim 2$~m$^2$~sec$^{-1}$, $L \sim 3 \times
10^6$~m, the condition becomes $B_* \gg 60$Re$_m^{1/2}$ nT.  For Re$_m
\sim 10^3$ the requirement becomes $B_* \gg 2$~$\mu$T.  $B_* \sim$
100~$\mu$T near the surface of the Sun: Assuming that the stellar
magnetic field decreases with the square of the distance (because of
interplanetary plasma), $St \gg 1$ only within 0.1~AU.  Thus
inhibition of convection could occur in the cores of the very closest
planets, although the increased temperature and conductivity of
surface rocks at such proximity to the star may block the stellar
field (see \S \ref{sec.other}).\\

\subsection{Other sources of magnetic fields \label{sec.other}}

In a planet is embedded in an ambient magnetic field from its star
that varies with orbital period $P$, eddy currents will generate an
induced field if a layer is sufficiently thick ($d$) and conducting.
The condition for a substantial induced field is \citep{Stevenson03}
\begin{equation}
d\sigma \gg \frac{1}{2\pi \mu_0 R_p}.
\end{equation}
For an Earth-size planet where $d$ is in km and $P$ in days, the
required conductivity $\sigma$ is $\gg 2 P d^{-1}
(R_p/R_{\oplus})^{-1}$~S~m$^{-1}$.  The conductivity of anhydrous
molten silicates is 0.1-1~S~m$^{-1}$ while that of molten carbonates
is three orders of magnitude higher \citep{Gaillard08}.  A planet with
a magma ocean only a few km deep on a planet with a very short-period
orbit such as CoRoT-7b (0.85~d) could have an induced field.\\

\begin{figure}
\noindent\includegraphics[width=3in]{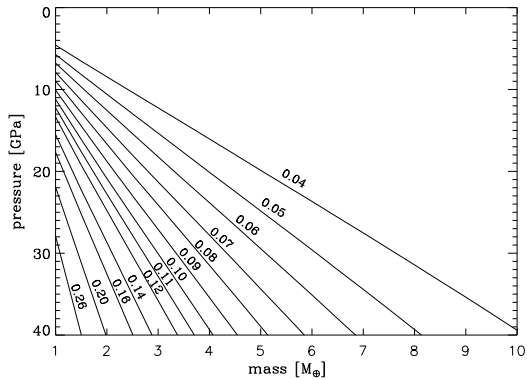}
\caption{Pressure at the base of a water (ice) mantle in solid planets
of varying mass and total water mass fraction.  Above 40~GPa, water
can be a sufficiently good ionic (proton) conductor ($\sim 2 \times
10^4$~S~m$^{-1}$) to produce an internal dynamo analogous to those
thought to operate in Uranus and Neptune.  There is also a minimum
temperature requirement that is not considered here.  The diameter of
GJ~1214b, $\sim$7~$M_{\oplus}$, 2.7~$R_{\oplus}$
\protect{\citep{Charbonneau09}}, indicates it has a massive volatile envelope
that could easily include the required water mantle.
\label{fig.ice}}
\end{figure}

Uranus and Neptune have magnetic fields that are thought to be
generated in a layer of high pressure water (ice) \citep{Cavazzoni99}.
The ionic conductivity of planetary ices is about $2 \times
10^3$~S~m$^{-1}$ at pressures above 40~Gpa, conditions that are
reached at 0.7-0.8 radii in these planets \citep{Stanley06}.  We added
liquid water and high-pressure ice (Ice VII) layers to the interior
model described in \S \ref{sec.interior} to calculate the pressure at
the base of an upper water/ice mantle as a function of total planet
mass and water mass fraction (Figure \ref{fig.ice}).  A basal pressure
of 40~Gpa is reached on planets of a few $M_{\oplus}$ and ice
fractions of $\sim$ 10\%.  Such an envelope of water could be easily
accomodated by the observed radius (2.7 $R_{\oplus}$) of GJ~1214b
\citep{Charbonneau09} but not by the mass and radius of CoRoT-7b
\citep{Valencia10}.\\

\subsection{Conclusions}

Our model of a planet consisting of a silicate mantle and pure Fe core
reproduces the gross interior structure and thermal history of the
Earth.  Using the magnetic field strength scaling of \citet{Aubert09},
the predicted evolution of the field strength is also consistent with
paleomagnetic inferences and the present value.  When applied to
Venus, with a stagnant lid, an elevated surface temperature, and
enhanced mantle viscosity, our model corrctly predicts the absence of
a dynamo at present.  We are thus confident in extrapolating the model
to rocky planets with approximately Earth-like composition but with
different mass, surface temperature, and/or mode of convection.\\

The model predicts that 1-2 \mearth~planets with plate tectonics will
sustain a dynamo for up to 7~Gyr by core cooling before an inner core
forms and the dynamo intensifies at a time that depends on surface
temperature, mass, and the relative size of the core.  Planets in this
mass range with stagnant lids will have dynamos only once an inner
core begins to grow.  Planets with higher surface temperatures can
more easily sustain a dynamo because the resulting higher mantle
temperatures lead to a lower viscosity and hence more rapid mantle
convection and core cooling.\\

Planets larger than about 2 \mearth~do not develop solid inner cores
because the adiabat is steeper than the solidus.  In such planets,
iron freezes first at or near the core-mantle boundary.  The latent
heat and any light elements released at the top of the core will not
be able to drive a long-lived dynamo.  In planets between 2-3 \mearth,
this leads to the premature shutdown of any dynamo within 10~Gyr.\\

In planets with masses $>$ 3 \mearth, iron never begins to freeze in
10~Gyr.  If these planets have plate tectonics, our model predicts
that they have relatively weak dynamos driven solely by core cooling
but which can last several Gyr, depending on mass and surface
temperature.  The mantles of planets with stagnant lids in this mass
range are hot enough to prevent core cooling and dynamo operation, but
they may also be partially molten.  This last scenario is not
addressed by our model.  Our model predicts that CoRoT-7b may have a
dynamo, but only if the planet has a mobile lid, and the strength of
the field at its surface will be less than that at of the present
Earth ($\sim 50 \mu$T).  Advances in our knowledge of the properties
of planetary materials at very high pressure will improve our ability
to make more quantitative predictions about the intensity and
longevity of core dynamos in massive rocky exoplanets.\\

\acknowledgments

EG is supported by NSF grant AST-0908419 and NASA grant NNX10AI90G,
CPC by NSF grant EAR-0914712, and MM and JH by NSF EAR-0855737.
Emilio Herrero-Bervera and Julien Aubert helped us with the IAGA
paleomagnetic database, and we acknowledge the useful comments by an
anonymous reviewer.

\clearpage




\clearpage







\clearpage
\renewcommand{\thefootnote}{\alph{footnote}}
\begin{table}
\label{tab.eosparams}
\caption{EOS Parameters}
\begin{tabular}{lllll}
Material & $\rho$ kg~m$^{-3}$ & $K_1$ [GPa] & $K_1'$ & Reference\\
\hline
Ice/water & 1000 & 2.2 & 4.0 & \citet{Sotin07}\\
HP ice (VII) & 1460 & 23.9 & 4.2 & \citet{Sotin07}\\
Mg-Olivine & 3222 & 128 & 4.3 & \citet{Sotin07}\\
Perovskite & 4260 & 266 & 3.9 &  \citet{Ahrens00}\\
Fe(l) & 7019 & 154\tablenotemark{a} & 4.66 & \citet{Anderson94}\\
Fe($\epsilon$) & 8315 & 165 & 4.97 & \citet{Vocadlo03}\\
\hline
\end{tabular}
\footnotetext[1]{Adjusted 40\% upwards from the \protect{\citet{Anderson94}} value.}
\end{table}

\begin{table}
\label{tab.thermoparams}
\caption{Thermodynamic Parameters}
\begin{tabular}{llcll}
Symbol & Parameter [units] & Material & Value & Reference\\ 
\hline
$c_p$ & specific heat capacity [J~kg$^{-1}$~K~$^{-1}$] & Fe & 850 & \citet{Wang02}\\
& `` '' & MgSiO$_3$-pv & 1250 & \citet{Akaogi93} \\
k & thermal conductivity [W~m$^{-1}$K$^{-1}$] & Fe & 30-100 & \citet{Stacey07} \\
& `` '' & MgSiO$_3$-pv & $\sim$6 & \citet{Goncharov09} \\
$\alpha$ & thermal expansivity [K$^{-1}$] & MgSiO$_3$-pv & $3 \times 10^{-5}$ & \citet{Anderson97} \\
$\gamma$ & Gr\"{u}neisen parameter & Fe & 1.5 & \citet{Labrosse07} \\
& `` '' & MgSiO$_3$-pv & 1.45 & \citet{Akaogi93}\\
$\Delta \rho /\rho$ & light element density deficit [\%] & Fe & 10 & \citet{Li03} \\
$\Delta$S & entropy of fusion [J~kg$^{-1}$K$^{-1}$] & Fe & 118 & \citet{Anderson97}\\
& `` '' & MgSiO$_3$-pv & 130 & \citet{Ito71} \\
$\lambda$ & magnetic diffusivity [m$^{2}$sec$^{-1}$] & Fe & 2 & \citet{Stevenson03}\\
$\kappa$ & thermal diffusivity [m$^{2}$sec$^{-1}$] & MgSiO$_3$-pv & $1 \times 10^{-6}$ & \citet{Turcotte02} \\
 & & Fe & $\sim 10^{-5}$ & calculated \\
$\sigma$ & electrical conductivity [S m$^{-1}$] & Fe & $5 \times 10^5$ & \citet{Bi02}\\
$\eta_*$ & reference viscosity [kg~m$^{-1}$sec$^{-1}$] & MgSiO$_3$-pv & $6 \times 10^{19}$ & tuning\\
$\tau_0$ & 1-bar melting point [K] & Fe & 1811 & \citet{Anderson97}\\
 &   liquidus at Earth CMB [K] & MgSiO$_3$-pv & 5400 & \citet{Stixrude09}\\
\hline
\end{tabular}
\end{table}




\end{document}